\documentstyle[twoside,12pt,epsf,amssymb,latexsym, amsmath]{article}

\begin{document}

\title{\bf Integration of nonlinear Partial Differential Equations 
 by using matrix algebraic systems}

\author{ Alexandre I. Zenchuk\\
Center of Nonlinear Studies of\\ L.D.Landau Institute 
for Theoretical Physics  \\
(International Institute of Nonlinear Science)\\
Kosygina 2, Moscow, Russia 119334\\
E-mail: zenchuk@itp.ac.ru\\}
\maketitle

\begin{abstract}
The paper  develops the method for construction of families of
particular solutions to some classes of nonlinear Partial Differential
Equations (PDE). Method
is based on the specific link between algebraic matrix equations
and PDE.
Admittable solutions involve arbitrary functions of either single
or several variables.
\end{abstract}

\section{Introduction}
Analysis  of nonlinear Partial Differential Equations (PDE) is
severe problem in mathematical physics.  Many different methods have
been developed for analytical investigation of nonlinear PDE
during last decades:
Inverse
Scattering Problem \cite{GGKM,ZSh1,ZSh2,ZM1,M,ZMNP,AKNS},
Sato theory \cite{OSTT,PM,P,Z4}, Hirota bilinear method \cite{H1,H2,HS,HS2}, 
Penlev\'e method \cite{WTC,W,EG}, 
$\bar{\partial}$-problem \cite{ZM,BM,K}, with some generalizations
\cite{DMZ,Z5,Z6,KZ,Z7},
quasi-classical version of the  $\bar{\partial}$-problem \cite{KMR,BKM,KMM}. 
A wide class of PDE  (so-called completely
integrable systems) has been studied better then others. 
Nevertheless, there are many methods which work in nonintegrable
case as well:  \cite{H1,H2,HS,HS2,WTC,W,EG}.  

We represent the method for construction of the families of
particular solutions to some classes of
multidimensional nonlinear PDE. This method is
 based on general properties of linear algebraic matrix
 equations. Essentially we 
  develop some ideas represented in the ref.\cite{OSTT}  and
recently in the ref. \cite{Z4}.  

General algorithm is discussed in the Sec. 2. We consider
systems, admitting solutions depending on arbitrary functions of
either single or several variables. Sec. 3 represents
some examples among which  are Kadomtsev-Petviashvili equation
(KP) 
and Devi-Stewartson equation (DS). 
Conclusions are given in the Sec.5.

\section{General results}

The algorithm represented in this section is based on the
fundamental properties of linear matrix algebraic equation, which
is written in the following form:
\begin{eqnarray}\label{MATR}
\Psi U = \Phi,
\end{eqnarray}
where $\Psi=\{\psi_{ij}\}$ is $N\times N$  matrix,
$U$ and  $\Phi$ are $N\times M$ matrices.
Let us recall these properties.

\begin{enumerate}
\item
 If $\Psi $ is nondegenerate matrix, i.e.
\begin{eqnarray}\label{DET}
\det \Psi\neq 0,
\end{eqnarray}
then equation (\ref{MATR}) has unique solution which can be written in the
following form:
\begin{eqnarray}\label{SOL}
U=\Psi^{-1} \Phi.
\end{eqnarray}
Only nondegenerate matrices $\Psi$ will be considered hereafter.
\item
({\it consequence of the previous property}) 
If $\Phi=0$ and condition (\ref{DET}) is held, then the 
equation (\ref{MATR}) has only the trivial solution
\begin{eqnarray}\label{SOLZ}
U\equiv 0.
\end{eqnarray}
\item
 ({\it superposition principle}) 
Consider the set of $K$ matrix equations with the same matrix $\Psi$:
\begin{eqnarray}\label{SUPER}
\Psi U_i = \Phi_i,\;\;i=1,\dots,K.
\end{eqnarray}
Then for any set of scalars $b_k$ ($k=1,\dots,K$),
 function $\tilde U=\sum_{k=1}^K b_k U_k$
is solution of the following matrix equation
\begin{eqnarray}\label{MATRSUPER}
\Psi \tilde U = \sum_{k=1}^K b_k\Phi_k.
\end{eqnarray}
\item
 ({\it consequence of properties 2 and 3})
If columns $\Phi_i$ are linearly dependent,
i.e there are scalars $a_k$, $k=1,\dots,K$, such that
\begin{eqnarray}\label{FUND1}
\sum_{k=1}^K a_k \Phi_k =0,\;\; 
\end{eqnarray} 
then 
\begin{eqnarray}\label{FUND2}
\sum_{k=1}^K a_k U_k  =0.
\end{eqnarray}
\end{enumerate}
Note that  analogous properties of linear integral equation have been
used in the 
classical dressing method based on the $\bar{\partial}$-problem
\cite{ZM,BM,K}.

We will use two sets of variables, which will be introduced in
the functions $\Phi$ and $\Psi$: $x=(x_1,\dots, x_Q)$ , $t=(t_1,t_2,\dots)$,
where $Q={\mbox{dim}}(x)$. 
The next statement follows from the above properties 
of the linear equations.

If there is transformation
$T$, which maps the nonhomogeneous equation (\ref{MATR}) into the
homogeneous
\begin{eqnarray}
\label{hom2}
\Psi \tilde U(U) =0, 
\end{eqnarray}
then
\begin{eqnarray}\label{hom3}
\tilde U(U)=0,
\end{eqnarray}
or
\begin{eqnarray}
\Phi=\Psi U\; \xrightarrow{T} \;0 =\Psi \tilde U(U) \Longrightarrow
 \tilde U(U) \equiv 0
\end{eqnarray}
If $\tilde U$ depends on
  $U$ and its derivatives, then   equation
({\ref{hom3})
 represents the
matrix PDE for $U$.

We will see that transformation $T$ is not unique.
One has the manifold  of transformations ${\Bbb T}$
 \begin{eqnarray}
 T_j\in {\Bbb T}:\;
\Phi=\Psi U  \; \xrightarrow{T_j} \;
0 =\Psi \tilde U_j(U)  \Longrightarrow\tilde U_j(U) \equiv 0
\end{eqnarray}
which is uniquely defined
by the equations  introducing  variables $x$ and $t$ in the
matrices $\Phi$ and $\Psi$.
Namely, 
\begin{eqnarray}\label{x}
\Psi_{x_k} = \Psi B_k + \Phi C_k,\;\;k=1,\dots,Q
\end{eqnarray}$
(B_k$ and $C_k$ are  $N\times N$ and $M\times N$ matrices
respectively) and
\begin{eqnarray}\label{t}
M_\alpha^n\Psi =0,\;\;\; M_\alpha^n\Phi = 0,\;\;
M_\alpha^n=\partial_{t_\alpha} + \partial^\alpha,\;\;
\sum_{k=1}^Q \alpha_k = n,
\end{eqnarray}
where $n$ is order of differential  operator.
 We use $Q$-dimensional vector subscript
$\alpha=(\alpha_1,\dots \alpha_Q)$ and notation $
\partial^\alpha=\prod_{j=1}^Q \partial_{x_j}^{\alpha_j}
$. Systems (\ref{x}) and (\ref{t}) should be compatible, which
leads to relations among matrices $B_j$ and $C_j$ (see the next
subsection).

 Each operator $M_\alpha^n$ defines transformation $T_\alpha\in
 {\Bbb T}$. In fact, let us apply  $M_\alpha^n$
to both sides of the eq. (\ref{MATR}) and use eqs. 
(\ref{t}):
\begin{eqnarray}\label{MPsi}
0=M_\alpha^n\Phi = (\partial_{t_\alpha}\Psi) U +\Psi
\partial_{t_\alpha} U +
\partial^\alpha (\Psi U) =
-(\partial_\alpha \Psi) U+\Psi
\partial_{t_\alpha} U +\\
\nonumber \sum_{j_1=0}^{\alpha_1} \dots
\sum_{j_Q=0}^{\alpha_Q}
\left(\prod_{k=1}^QC_{\alpha_k}^{j_k}\right)
\left(\prod_{n=1}^Q\partial_{x_n}^{j_n}\Psi\right)
\left(\prod_{m=1}^Q\partial_{x_m}^{\alpha_m-j_m} U\right),
\end{eqnarray} 
$\displaystyle C_n^k=\frac{n!}{k! (n-k)!}$ are binomial
coefficients.
 Due to the eqs.(\ref{MATR}) and (\ref{x}) , one has the  
 following 
 expression for $ \partial_{x_j}\Psi$:
 \begin{eqnarray}
  \partial_{x_j}\Psi=
 \Psi V_{e^j} ,\;\; \;\;
 V_{e^j}=B_j + U C_j,\;\;
 e^j=(\underbrace{0,\dots,0}_{j-1},1,
 \underbrace{0,\dots,0}_{Q-j} ).
 \end{eqnarray}
Analogously  we can write 
 \begin{eqnarray}\label{dpsi}
 \partial^\beta \Psi =  \Psi V_\beta,
 \end{eqnarray}
 where $V_{\beta}$ can be found recursively. In fact, one has
 \begin{eqnarray}
\partial^\beta \Psi = \partial_{x_j}
\partial_{(\beta_1,\dots,\beta_j-1,\dots,\beta_Q)}
\Psi=
\partial_{x_j}( \Psi
V_{(\beta_1,\dots,\beta_j-1,\dots,\beta_Q)})=\\
\nonumber
(\partial_{x_j} \Psi)V_{(\beta_1,\dots,\beta_j-1,\dots,\beta_Q)}
+
\Psi (\partial_{x_j}
V_{(\beta_1,\dots,\beta_j-1,\dots,\beta_Q)})=\\
\nonumber
\Psi \left(V_{e^j} V_{(\beta_1,\dots,\beta_j-1,\dots,\beta_Q)}
+\partial_{x_j}
V_{(\beta_1,\dots,\beta_j-1,\dots,\beta_Q)}\right)
 \end{eqnarray}
 i.e.
 \begin{eqnarray}
 V_\beta=V_{e^j} V_{(\beta_1,\dots,\beta_j-1,\dots,\beta_Q)}
+\partial_{x_j}
V_{(\beta_1,\dots,\beta_j-1,\dots,\beta_Q)},\\\nonumber
V_0=V_{(\underbrace{0,\dots,0}_{Q})}=I_{N,N}.
 \end{eqnarray}
 Owing to the eq.(\ref{dpsi}) we can replace factor 
 $\left(\prod_{n=1}^Q\partial_{x_n}^{j_n}\Psi\right)$
 in (\ref{MPsi}) with $\Psi V_{(j_1,\dots, j_Q)}$, which results in
 \begin{eqnarray}\label{nl1}
 0=-(\partial_\alpha \Psi) U+\Psi
\partial_{t_\alpha} U +\\\nonumber
\Psi \sum_{j_1=1}^{\alpha_1} \dots
\sum_{j_Q=1}^{\alpha_Q}
\left(\prod_{k=1}^QC_{\alpha_k}^{j_k}\right)
 V_{(j_1,\dots, j_Q)} \prod_{m=1}^Q  
\partial_{x_m}^{\alpha_m-j_m}
 U=
\Psi \tilde U_\alpha,\\\label{nl2}
\tilde U_\alpha= \partial_{t_\alpha} U +{\sum_{j_1,\dots
j_Q}}'
\left(\prod_{k=1}^QC_{\alpha_k}^{j_k}\right)
 V_{(j_1,\dots, j_Q)} \prod_{m=1}^Q  
\partial_{x_m}^{\alpha_m-j_m} U,
 \end{eqnarray}
 where $\displaystyle{\sum\limits_{j_1,\dots,
j_Q}}'$ means the sum over all values $j_1,\dots,
j_Q$ such that $(\sum_{m=1}^Q j_m)< n$
 The associated system of nonlinear PDE  has the form  (see
 (\ref{hom3}))
 \begin{eqnarray}\label{gennl}
\partial_{t_\alpha} U +{\sum_{j_1,\dots
j_Q}}'
\left(\prod_{k=1}^QC_{\alpha_k}^{j_k}\right)
 V_{(j_1,\dots, j_Q)} \prod_{m=1}^Q  
\partial_{x_m}^{\alpha_m-j_m} U=0
 \end{eqnarray}
 
 \subsection{Compatibility of the system (\ref{x}) and (\ref{t})}
 
 Studying the compatibility condition for the system (\ref{x}) we
 concentrate on two particular cases
 \begin{enumerate}
 \item
 compatibility condition produces the additional differential
 equation 
 for the function $U$ and $Q-1$ differential equations for the
 function $\varphi$. In this case  $\phi$ (and $U$) involve $N$ 
  arbitrary
scalar  functions 
  of single variable.
  \item
   compatibility condition produces only the differential
 equation 
 for the function $U$. In this case $\phi$ (and $U$) involve
 $N\times M$  arbitrary
 scalar  functions
 of $Q$ variables.
 \end{enumerate}
 
 Intermediate case (i.e. $\varphi$ and $U$ depend on  arbitrary
 functions of $P$ ($1<P<Q$) variables)  is also possible, 
 but it is not regarded in this
 paper.
  
 \subsubsection{Matrix  $U$ depends on  $N$ arbitrary
 scalar functions of single variable}

 We start the analysis with splitting the 
  eq. (\ref{x})  into  two equations
 by using the following structure of the matrices:
 \begin{eqnarray}\label{BCC1}
 \Psi=\left[
 \begin{array}{c|c}
 \varphi & \chi
 \end{array}
 \right],\;\; 
  B_j=\left[
 \begin{array}{c|c}
 b_j & B_{1j}\cr
 \hline
 0_{N-M,M}&B_{2j}
 \end{array}\right],\\\label{BCC2}
 C_j=\left[\begin{array}{c|c}
 C_{0j}&C_{1j}
 \end{array}\right],\;\;C_{01}=I_M,\;\;j=1,\dots, Q
 \end{eqnarray} 
 where
 $\varphi$ is $N\times M$, $\chi$ is $N\times (N-M)$;
 $b_{j}$, $C_{0j}$ are $M\times M$,
 $B_{1j}$ and $C_{1j}$ are $M\times (N-M)$;
 $B_{2j}$ are  $(N-M)\times (N-M)$ matrices for all
 $j=1,\dots,Q$; hereafter $0_{A,B}$ with arbitrary $A$ and $B$ 
 is $A\times B$ matrix of zeros; 
  $I_M$ is $M\times M$  identity
 matrix.
Having matrices with given structure one can split the system (\ref{x})
as follows:
\begin{eqnarray}\label{varphi1}
\varphi_{x_j} = \varphi b_j + \Phi C_{0j},\\\label{chi1}
\chi_{x_j} = \chi B_{2j} + \varphi B_{1j} + \Phi C_{1j}
\end{eqnarray}
 The eq.(\ref{varphi1}) with $j=1$ defines $\Phi$ in terms of 
 $\varphi$ (remember, that $C_{01}=I_{M}$):
 \begin{eqnarray}
 \Phi = \varphi_{x_1} - \varphi b_1,
  \end{eqnarray}
  Substituting this expression into the eq.(\ref{varphi1}) with $j>1$
  one gets the overdetermined system of equations for $\varphi$  :
  \begin{eqnarray}\label{varphij}
 \varphi_{x_j} = \varphi_{x_1} C_{0j} + \varphi (b_j-b_1
 C_{0j}),\;\;j>1.
 \end{eqnarray}
 Compatibility of this system with different $j$ 
 leads to the following relations among matrices:
 \begin{eqnarray}\label{BC1}
 [C_{0k},C_{0j}]=0, \\\nonumber
 [ b_j,C_{0k} ]-[b_k,C_{0j}]+C_{0k} b_1 C_{0j} - C_{0j} b_1
 C_{0k}=0, \\\nonumber
 [b_j-b_1 C_{0j}, b_k - b_1 C_{0k}]=0.
 \end{eqnarray}
 From another point of view, eqs. (\ref{MATR}), (\ref{x})
 and (\ref{varphi1}) 
  lead to the additional differential equation for
 $U$:
  \begin{eqnarray}\label{comp1}
E_M [b_i,b_j] + U C_{0i} b_j + U_{x_i} C_{0j} + B_i U C_{0j} + U C_i U C_{0j} 
=\\\nonumber
U C_{0j} b_i +  U_{x_j} C_{0i} + B_j U C_{0i} + U C_j U C_{0i},
 \end{eqnarray}
where $E_M$ is matrix,composed of first $M$ column of the 
$N$-dimensional identity matrix $I_N$.  So, 
eq. (\ref{varphi1}) describes one more transformation 
 $T^{\mbox{comp}}\in {\Bbb  T}$.

 The system (\ref{chi1}) is overdetermined system defining $\chi$ in 
 terms of $\varphi$. 
 Its compatibility condition 
  leads to the following relations 
 \begin{eqnarray}\label{BC2}
 A_3 \chi + A_2 \varphi_{x_1x_1} + A_1 \varphi_{x_1} + A_0
 \varphi =0,\\\nonumber
 A_3= [B_{2i},B_{2j}] \equiv0\\\nonumber
 A_2=C_{0i} C_{1j} - C_{0j} C_{1i} \equiv 0,\\\nonumber
 A_n=A_n(C_{0i},C_{0j},C_{1i},C_{1j},B_j) \equiv 0,\;\;n=0,1.
 \end{eqnarray}
 We don't represent expressions for $A_0$ and $A_1$, because they
 are too complicated.
 But in the particular case, when
 $b_j=0_{M,M}$ and $C_{1j}=0_{M,N-M}$ the above system has simpler form
 \begin{eqnarray}\label{BC3}
 [B_{2i},B_{2j}] = 0\\\nonumber
C_{0i} C_{1j} - C_{0j} C_{1i} =0,\\\nonumber
C_{1i} B_{2j} + C_{0i} B_{1j} - C_{1j} B_{2j} - C_{0j} B_{1i}=0\\\nonumber
B_{1i}B_{2j} - B_{1j} B_{2i} = 0.
 \end{eqnarray}

To satisfy both systems
(\ref{x}) and (\ref{t}) $\varphi$ should  be written in the form:
\begin{eqnarray}
\varphi=\int _{-\infty}^\infty
c(k_1) \exp\left[\sum_{i=1}^Q k_i x_i + \sum_{\alpha} \omega_\alpha
t_\alpha\right] dk_1,
\end{eqnarray}
where $c(k_1)$ ($N\times M$ matrix function of
argument)  and  $k_j$ satisfy the dispersion relation associated with
eq.(\ref{varphij}), $\omega_\alpha$ satisfy the dispersion
relation for eq.(\ref{t}). One can see that the above expression
for $\varphi$  (and $U$ due to the eq. (\ref{SOL}))
involves $N$ arbitrary functions of single
variable.
Nonlinear equations generated by the operators $M_\alpha^n$ have
the  form (\ref{gennl})

\subsubsection{Matrix $U$ 
depending on $N\times M$  arbitrary scalar functions of $Q$
variables}

In this section we consider 
matrix equation (\ref{MATR}) in the form
\begin{eqnarray}
\Phi=\Psi U,\;\;
\Phi=\left[
\begin{array}{ccc}
\Phi_1 & \cdots & \Phi_P
\end{array}
\right],\;\;
U=\left[
\begin{array}{ccc}
U_1 & \cdots & U_P
\end{array}
\right],\;\;{\mbox{or}}\\\label{Phik}
\Phi_k=\Psi U_k,\;\;k=1,\dots,P,
\end{eqnarray}
where $\Phi_j$ and $U_j$ are $N\times M$, 
$\Psi$ is $N\times N$
matrices.
Let matrix $C_k$ be of the form:
\begin{eqnarray}
C_k=\left[
\begin{array}{c}
0_{M(k-1),N}\cr
\tilde C_k \cr
0_{N-M k,N}
\end{array}
\right],
\end{eqnarray}
where $\tilde C_k$ is $M\times N$ matrix.
So that eq.(\ref{x}) can be written as follows:
\begin{eqnarray}\label{x2}
\Psi_{x_k} = \Psi B_k + \Phi_k \tilde C_k,\;\;k=1,\dots,Q,
\end{eqnarray}
variables $t_n$ are introduced by the same eq.(\ref{t}).
Assume the following structure of matrices:
\begin{eqnarray}
\Psi=
\left[
\begin{array}{c|c}
\varphi & \chi
\end{array}
\right],\;\;
\tilde C_k=\left[
\begin{array}{c|c}
I_M & C_{1k}
\end{array}
\right],\;\;
B_k=\left[
 \begin{array}{c|c}
 b_k & B_{1k}\cr
 \hline
 0_{M,N-M}&B_{2k}
 \end{array}\right],
\end{eqnarray}
where
 $b_{k}$, are $M\times M$ matrices,
 $C_{1k}$ and $B_{1k}$ are $M\times (N-M)$ matrices,
 $B_{2k}$ are  $(N-M)\times (N-M)$ matrices for all
 $k=1,\dots,Q$.
 Now we can split the eq.(\ref{x2}) into the following pair of
 equations
 \begin{eqnarray}\label{Phi2}
 \varphi_{x_k} = \varphi b_k +\Phi_k ,\\\label{chi2}
 \chi_{x_k} = \chi B_{2k} + \varphi B_{1k} + \Phi_k C_{1k}.
 \end{eqnarray}
Equation (\ref{Phi2}) defines function $\Phi_k$ in terms of
$\varphi$. 
At the same time, $\Phi_k$ can be eliminated from the
eq.(\ref{Phi2}) due to the eq.(\ref{Phik}), so that
eq.(\ref{Phi2}) represents overdetermined system for $\varphi$
with compatibility condition (compare with eq.(\ref{comp1}))
\begin{eqnarray}
E_M [b_i, b_j] + U_{x_i}^j + U^ib_j +
B_i U^j + U^i \tilde C_i U^j=\\\nonumber
 U_{x_j}^i + U^jb_i +
B_j U^i + U^j \tilde C_j U^i,
\end{eqnarray}
So, additional transformation 
$T^{\mbox{comp}}\in \Bbb T$ is based on the system (\ref{Phi2}).

Compatibility of the system (\ref{chi2}) gives relations among
matrices $B_j$ and  $C_j$: for all $j$ and $k$
\begin{eqnarray}\label{comp00}
C_{1j}-C_{1k} &=&0,\;\;\\\nonumber
[B_{2j},B_{2k}]&=&0,\\\nonumber
B_{1j}-b_j C_{1j} + C_{1j} B_{2j}&=&0.
\end{eqnarray}
Expression for $\varphi$ satisfying both systems (\ref{x}) and
(\ref{t}) is following:
\begin{eqnarray}\label{varphi2}
\varphi=\int _{-\infty}^\infty
c(k_1,\dots,k_Q) \exp\left[\sum_{i=1}^Q k_i x_i + 
\sum_{\alpha} \omega_\alpha
t_\alpha\right] dk_1\dots dk_Q,
\end{eqnarray}
where $c$ is $N\times M$ arbitrary matrix function of arguments;
$\omega_\alpha$ satisfy the dispersion relations for (\ref{t}).
Due to the formula (\ref{SOL})  and (\ref{varphi2}), $U$ depends
on $N\times M$ arbitrary functions of $Q$ variables.
Nonlinear equations generated by the operators $M_\alpha^n$ keep
the  form (\ref{gennl})

\section{Examples}

We consider two hierarchies  of PDE, admitting solutions depending on
$N$ functions of single variables (see subsection 2.1.1). Inside
of them are classical  KP and DS hierarchies.  
It is important to note that
$N$ can be arbitrary integer for later, which is not true  in
general. 

\subsection{KP hierarchy}

Let $M=1$, $Q=1$, $M^n=\partial_{t_n} + \partial_x^n$,
$n=2,3,\dots$.
Operators $M^n$ generate the following hierarchy (see
eq.(\ref{gennl})):
\begin{eqnarray}\label{kp}
U_{t_n} + \sum_{i=1}^n C_n^i V_{n-i} \partial_x^i U =0
,\\\nonumber
V_0=I_{N},;\; V_1= B+U C,\;\; V_n=V_1 V_{n-1} +\partial_x V_{n-1},
\end{eqnarray}
where 
 $B\equiv B_1$ and $C\equiv C_1$.
In this case one has only one eq.(\ref{x}) and no compatibility
condition for it, i.e. matrices $B$ and $C$ are {\it arbitrary}
matrices having structure (\ref{BCC1}) and (\ref{BCC2}). 
If 
\begin{eqnarray}
B=\left[
\begin{array}{c|cccc}
0&1&0&0&\dots\cr
\hline
0&0&1&0&\dots\cr
0&0&0&1&\dots\cr
\vdots&\vdots&\vdots&\vdots&\vdots
\end{array}
\right],\;\;C=\left[\begin{array}{c|ccc}
1&0&0&\dots\end{array}\right],\;\;
U=\left[\begin{array}{c}
u_1\cr\vdots\cr u_N\end{array}\right],
\end{eqnarray}
$N$ is
arbitrary, then system (\ref{kp}) represent KP hierarchy \cite{OSTT}, .  
KP can be written for the function $u\equiv {u_1}_x$ using
eq.(\ref{kp}) with $n=2,3$:
\begin{eqnarray}
u_t +\frac{1}{4} u_{xxx} + 
\frac{3}{4} \partial^{-1}_x u_{yy} + 3 u u_x = 0
\end{eqnarray}

\subsection{DS hierarchy}
Let $M=2$, $Q=2$, $x_1=x$, $x_2=y$.
Consider operators $M^n=\partial_{t_n} + \partial_x^n$.
This operators generate the same hierarchy (\ref{kp}) with
appropriate dimensions for matrix $U$,
\begin{eqnarray}
V_1=B_1 + U C_1,
\end{eqnarray}
and additional nonlinear equation (\ref{comp1}):
\begin{eqnarray}\label{ds}
 U_{x} C_{02} + B_1 U C_{02} + U C_1 U C_{02} =
  U_{y} C_{01} + B_2 U C_{01} + U C_2 U C_{01}
 \end{eqnarray}
 with relations among matrices $B_j$ and $C_j$ given by 
 (\ref{BC1}) and (\ref{BC2}) (or (\ref{BC3})).
 DS hierarchy corresponds to the following choice of matrices
 $B_j$ and $C_j$:
\begin{eqnarray}
B_1=\left[
\begin{array}{cc|cccc}
0&0&1&0&0&\dots\cr
0&0&0&1&0&\dots\cr
\hline
0&0&0&0&1&\dots\cr
\vdots&\vdots&\vdots&\vdots&\vdots&\vdots
\end{array}
\right],\;\;
B_2=\left[
\begin{array}{cc|ccccc}
0&0&0&1&0&0&\dots\cr
0&0&1&0&0&0&\dots\cr
\hline
0&0&0&0&0&1&\dots\cr
0&0&0&0&1&0&\dots\cr
\vdots&\vdots&\vdots&\vdots&\vdots&\vdots&\vdots
\end{array}
\right],\;\;\\\nonumber
 C_1=\left[\begin{array}{cc|cc}
1&0&0&\dots\cr
0&1&0&\dots\end{array}\right],\;\;
 C_2=\left[\begin{array}{cc|cc}
0&1&0&\dots\cr
1&0&0&\dots\end{array}\right],\;\;
U=\left[\begin{array}{cc}
u_1&v_1\cr\vdots&\vdots\cr u_N&v_n\end{array}\right],
\end{eqnarray}
$N$ is arbitrary.
It is simply to check that conditions (\ref{BC1},\ref{BC3}) 
are satisfied.
Let us consider  the eq.(\ref{kp}) with $n=2$
together with eq.(\ref{ds}). The following    
change of variables
\begin{eqnarray}
u_1=\frac{1}{4} (r-q+2 w_x),\;\;u_2=\frac{1}{4} (r+q+2
w_y),\;\;\\\nonumber
v_1=\frac{1}{4} (-r-q+2 w_y),\;\;v_2=\frac{1}{4} (-r+q+2
w_x).\;\;
\end{eqnarray}
transforms them into 
\begin{eqnarray}
 r_{t_2} - r_{xy} -2 r w_{xy} =0,\;\;\;
 q_{t_2} + q_{xy} +2 q w_{xy} =0 ,\;\;\;
 w_{xx} - w_{yy} = qr ,
\end{eqnarray}
which results in  DS after reduction
$r=\psi$, $q=\bar\psi$, $t_2 = i t$, where $i^2=-1$, bar means complex
conjugated value 

\section{Conclusions}

The suggested version of the dressing method is one more form 
of
representation of the systems of nonlinear PDE,
which admit an infinite number of commuting flows, generated by
the operators $M_\alpha^n$. Among them are classical completely
integrable systems of equations. The feature of these system is
that their solutions  admit {\it any} number of arbitrary scalar 
 functions of single variable ($N$ is arbitrary for them). 
 One can see that there is another type of systems, which admit an
 infinite number of commuting flows, while $N$ is given. We don't
 know interesting examples for this case.
 We also have not found the systems,
 whose solutions  would admit any number of arbitrary scalar functions of
 several variables: for all nontrivial examples this number has to
 be fixed.
 These aspects will be studied later.

The work 
is supported by RFBR grants 01-01-00929 and 00-15-96007.

\end{document}